\documentclass[
 reprint,
superscriptaddress,
nofootinbib,
 amsmath,amssymb,
aps,
prx,
]{revtex4-2}

\usepackage{graphicx}% Include figure files
\usepackage{dcolumn}% Align table columns on decimal point
\usepackage{bm}% bold math
\usepackage[T1]{fontenc}
\usepackage{comment}
\usepackage{xcolor}
\usepackage{siunitx}
\begin{document}

\preprint{APS/123-QED}
 
\title{Scalable Atomic Arrays for Spin-Based Quantum Computers in Silicon}

\author{Alexander M. Jakob$^*$$^\dagger$}
\author{Simon G. Robson$^\dagger$}
\affiliation{School of Physics, University of Melbourne, Parkville, 3010, VIC, Australia}
\affiliation{ARC Centre for Quantum Computation and Communication Technology (CQC$^2$T)}
\email{alexander.jakob@unimelb.edu.au, d.jamieson@unimelb.edu.au}

\author{Hannes R. Firgau}
\affiliation{School of Electrical Engineering and Telecommunications, UNSW Sydney, 2052, NSW, Australia}
\affiliation{ARC Centre for Quantum Computation and Communication Technology (CQC$^2$T)}
\author{Vincent Mourik}
%\affiliation{JARA-FIT Institute for Quantum Information, Forschungszentrum J{\"u}lich, 52056, North Rhine Westphalia, Germany}
\affiliation{School of Electrical Engineering and Telecommunications, UNSW Sydney, 2052, NSW, Australia}
\affiliation{ARC Centre for Quantum Computation and Communication Technology (CQC$^2$T)}
\author{Vivien Schmitt}
%\affiliation{}
\affiliation{School of Electrical Engineering and Telecommunications, UNSW Sydney, 2052, NSW, Australia}
\affiliation{ARC Centre for Quantum Computation and Communication Technology (CQC$^2$T)}
\author{Danielle Holmes}
\affiliation{School of Electrical Engineering and Telecommunications, UNSW Sydney, 2052, NSW, Australia}
\affiliation{ARC Centre for Quantum Computation and Communication Technology (CQC$^2$T)}

\author{Matthias Posselt}
\affiliation{Helmholtz-Zentrum Dresden-Rossendorf (HZDR), Dresden, 01328, Saxony, Germany}

\author{Edwin L.H. Mayes}
\affiliation{RMIT Microscopy and Microanalysis Facility, RMIT University, Melbourne, Victoria 3001, Australia}

\author{Daniel Spemann}
\affiliation{Leibniz-Institut f{\"u}r Oberfl{\"a}chenmodifizierung e.V., Leipzig, 04318, Saxony, Germany}

\author{Andrea Morello}
\affiliation{School of Electrical Engineering and Telecommunications, UNSW Sydney, 2052, NSW, Australia}
\affiliation{ARC Centre for Quantum Computation and Communication Technology (CQC$^2$T)}

\author{David N. Jamieson$^*$}%
\affiliation{School of Physics, University of Melbourne, Parkville, 3010, VIC, Australia}
\affiliation{ARC Centre for Quantum Computation and Communication Technology (CQC$^2$T)}
\email{d.jamieson@unimelb.edu.au}

\date{\today}

\begin{abstract}
\noindent 
Semiconductor spin qubits combine excellent quantum performance with the prospect of manufacturing quantum devices using industry-standard metal-oxide-semiconductor (MOS) processes. This applies also to ion-implanted donor spins, which further afford exceptional coherence times and large Hilbert space dimension in their nuclear spin. Here we demonstrate and integrate multiple strategies to manufacture scale-up donor-based quantum computers. We use $^{31}$PF$_{2}$ molecule implants to triple the placement certainty compared to $^{31}$P ions, while attaining 99.99\% confidence in detecting the implant. Similar confidence is retained by implanting heavier atoms such as $^{123}$Sb and $^{209}$Bi, which represent high-dimensional qudits for quantum information processing, while Sb$_2$ molecules enable deterministic formation of closely-spaced qudits. We demonstrate the deterministic formation of regular arrays of donor atoms with 300~nm spacing, using step-and-repeat implantation through a nano aperture. These methods cover the full gamut of technological requirements for the construction of donor-based quantum computers in silicon.   
\end{abstract}

\maketitle
\def\thefootnote{$^\dagger$}\footnotetext{These authors contributed equally to this work}

%\tableofcontents

%\subsection*{Main}
Among the practical implementations of solid-state spin qubits \cite{zwanenburg2013,chatterjee2021semiconductor,wolfowicz2021quantum,burkard2023semiconductor}, ion-implanted donor systems \cite{morello2020donor} occupy a special place by virtue of retaining compatibility with industry-standard metal-oxide-semiconductor (MOS) manufacturing process \cite{pillarisetty2018qubit,gonzalez2021scaling}, combined with exceptionally long coherence times (exceeding 30 seconds for the $^{31}$P nuclear spin \cite{muhonen2014}) and one- and two-qubit gate fidelities exceeding 99\% \cite{Muhonen2015,Madzik2022}. The latter breakthrough means that it becomes possible to envisage the construction of two-dimensional (2D) arrays of donor spin qubits to be operated as logical qubits, for example within the surface code architecture \cite{Fowler2012, Reiher7555}. Such architecture also affords a $^\sim$5-10\% fraction of defective site in the arrays \cite{Nagayama2017, Auger2017}. The efficiency of the error correcting code and the tolerance against physical errors could be further improved by implementing single-step parity check gates \cite{ustun2023single}, which are native to donor nuclear spins sharing a common electron \cite{Madzik2022}.

This work presents a complete suite of deterministic material doping methods, suitable for the construction of large-scale donor arrays for quantum information processing in silicon. We define two basic types of scale-up strategies. ``Outward'' scaling is the standard method \cite{vandersypen2017interfacing} whereby physical two-level systems (in this case, electrons or nuclei possessing a spin-1/2) are placed in regular arrays and subjected to multi-qubit logic operations and quantum error correction protocols. We envisage 2D arrays of atoms spaced by $\approx 300$~nm, as appropriate for the operation of dipolarly-coupled `flip-flop' qubits \cite{Tosi2017,Savytskyy2022}. ``Inward'' scaling is the use of high-dimensional physical constituents (in this case, nuclei with spin larger than 1/2) to enlarge the Hilbert space dimension available for quantum information encoding, without the need to connect distinct physical objects. Such $d$-dimensional (with $d>2$) systems, known as qudits \cite{wang2020qudits}, form a rich quantum information platform for which high-spin donor nuclei \cite{morley2010initialization,wolfowicz2013,asaad2020,fernandez2023navigating} constitute one of only few solid state examples \cite{godfrin2018generalized,hendriks2022coherent,adambukulam2023hyperfine}.

We address the materials and manufacturing challenges of outward scaling \cite{Jamieson2005,vanDonkelaar2015,Pacheco2017,Raecke2019,Cassidy2020,Jakob2022,Robson2022} by employing $^{31}$PF$_2$ molecule ions to enable high-precision near-surface donor placement. Next, we demonstrate high-precision near-surface implantation of the heavy donors $^{123}$Sb ($I=7/2$) and $^{209}$Bi ($I=9/2$) which embody inward scaling platforms. Finally, by employing Sb$_2$ molecule ions, attributes of both strategies may be combined and further advanced. All strategies require donor atoms to be placed $^\sim12\pm6$~nm below a pre-fabricated surface gate oxide of typically $^\sim3-8$~nm thickness \cite{Madzik2021}. This requirement constrains the ion implantation energy to sub-10~keV for $^{31}$P donors and sub-20~keV for $^{123}$Sb and $^{209}$Bi, and we demonstrate deterministic doping techniques capable of detecting such low-energy implants with confidence close to or above 99.99\%.

%******************************************************************************************

\subsection*{Outward scaling: Background \& challenges}
\begin{figure}[t!]
\centering
\includegraphics[width=1.0\linewidth]{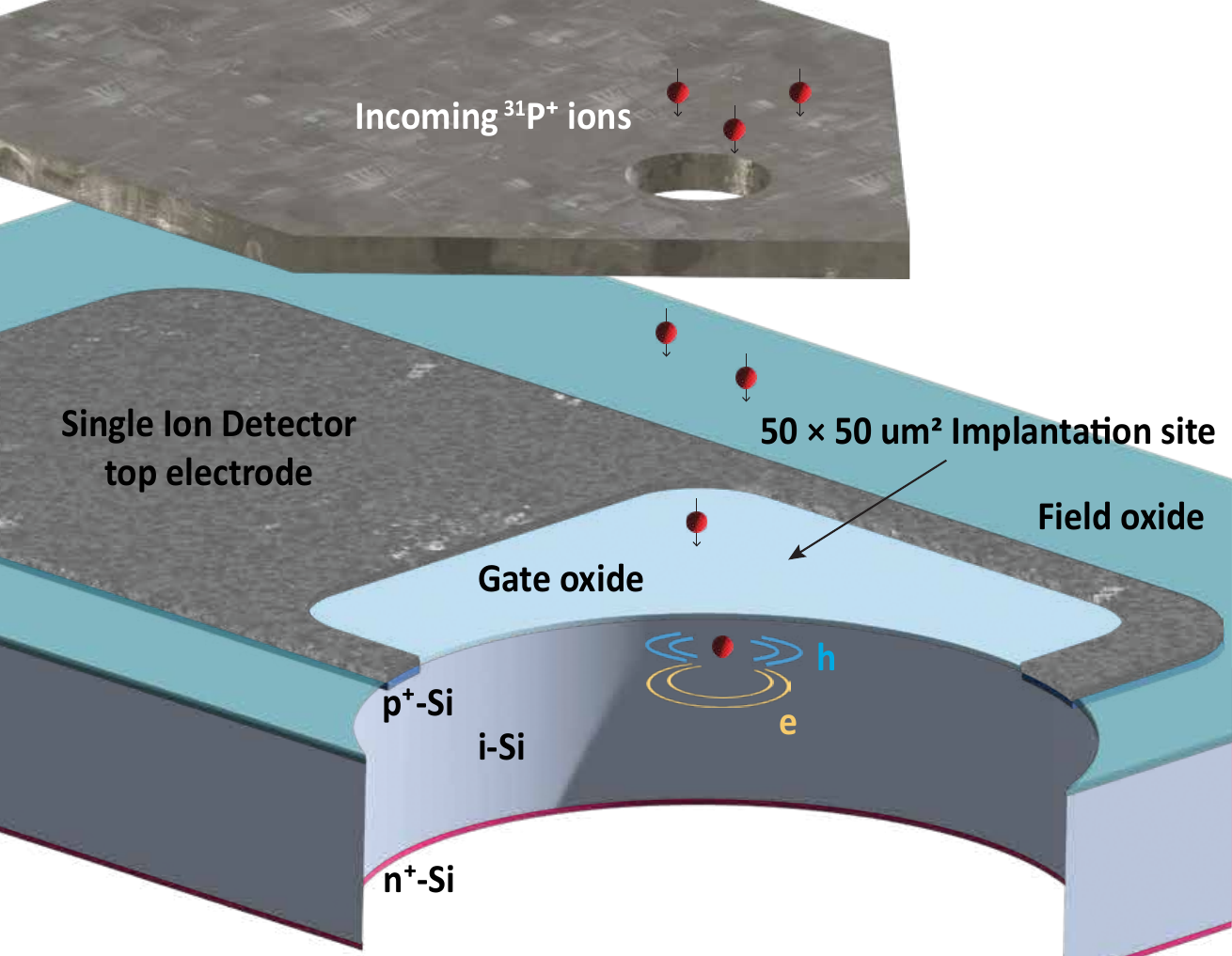}
\caption{Ion implantation configuration: An atomic-force microscope (AFM) cantilever with an aperture dwells over an implantation site on the silicon substrate configured with biased, charge sensitive detector electrodes. The substrate is passivated with a 5~nm thin gate oxide.  Implanted ions dissipate kinetic energy and create free electron-hole pairs that induce a signal at the detector electrodes. The signal amplitude is proportional to the number of electron-hole pairs and can be used to trigger a step-and-repeat sequence for the deterministic engineering of donor arrays. For the spectra presented in Figs. \ref{fig:Fig2} and \ref{fig:Fig3} the cantilever aperture had a diameter of 16 $\mu$m. For the higher placement precision required in Fig. \ref{fig:Fig5} the diameter was 45~nm.}
\label{fig:Fig1}
\end{figure}
The deterministic single ion implantation method utilised here employs the silicon substrate itself to register single ion implant events (see Fig. \ref{fig:Fig1}). The substrate is equipped with on-chip surface electrodes forming a p-i-n junction relative to the back contact and operated in reverse bias as a detector. Free electron-hole pairs ('charge carriers') are induced by the kinetic energy fraction of each ion which dissipates via electronic stopping, i.e. the stopping mechanism causing ionization in the silicon crystal \cite{Ziegler1985}. The resulting free carriers drift in the internal electric field and are collected at the electrodes, inducing a signal pulse with amplitude $q$ proportional to the number of ion-induced charge carriers \cite{Breese2007}. The energy needed to create one electron-hole pair in silicon ($\varepsilon=3.6$~eV) is used to give the signal amplitude $q$ in units of the equivalent energy. This signal is used to trigger a routine for constructing arrays of individual dopants. The signal amplitude also contains information about the fraction of kinetic energy dissipated to nuclear stopping, i.e the mechanism that leads to phonon excitation and lattice defects from energy transfer to target nuclei. If impacting ions are captured in crystal channels along low-index lattice axes, the nuclear stopping is reduced in favour of electronic stopping. This results in a much deeper implantation in the substrate, accompanied by a higher signal $q$. Therefore, events with abnormally high $q$ can be associated with donors that stopped too deep in the substrate, and trigger appropriate mitigation strategies.

The detection confidence for a single ion impact signal is limited by the probability that it cannot be distinguished from a noise-induced signal. In turn, the confidence $\varXi$ for an ensemble $s(q)=\{q_1,...q_i\}$ of ion impact signal events (i.e. signal 'spectrum' or 'histogram'), can be obtained from its overlap with the detector noise spectrum $n(q)$. This confidence can be optimised with a signal noise discriminator threshold (NDT) that discards all signals $q\leq q_\mathrm{NDT}$ \cite{Jakob2022}.

\begin{figure}[t!]
\centering
\includegraphics[width=1.0\linewidth]{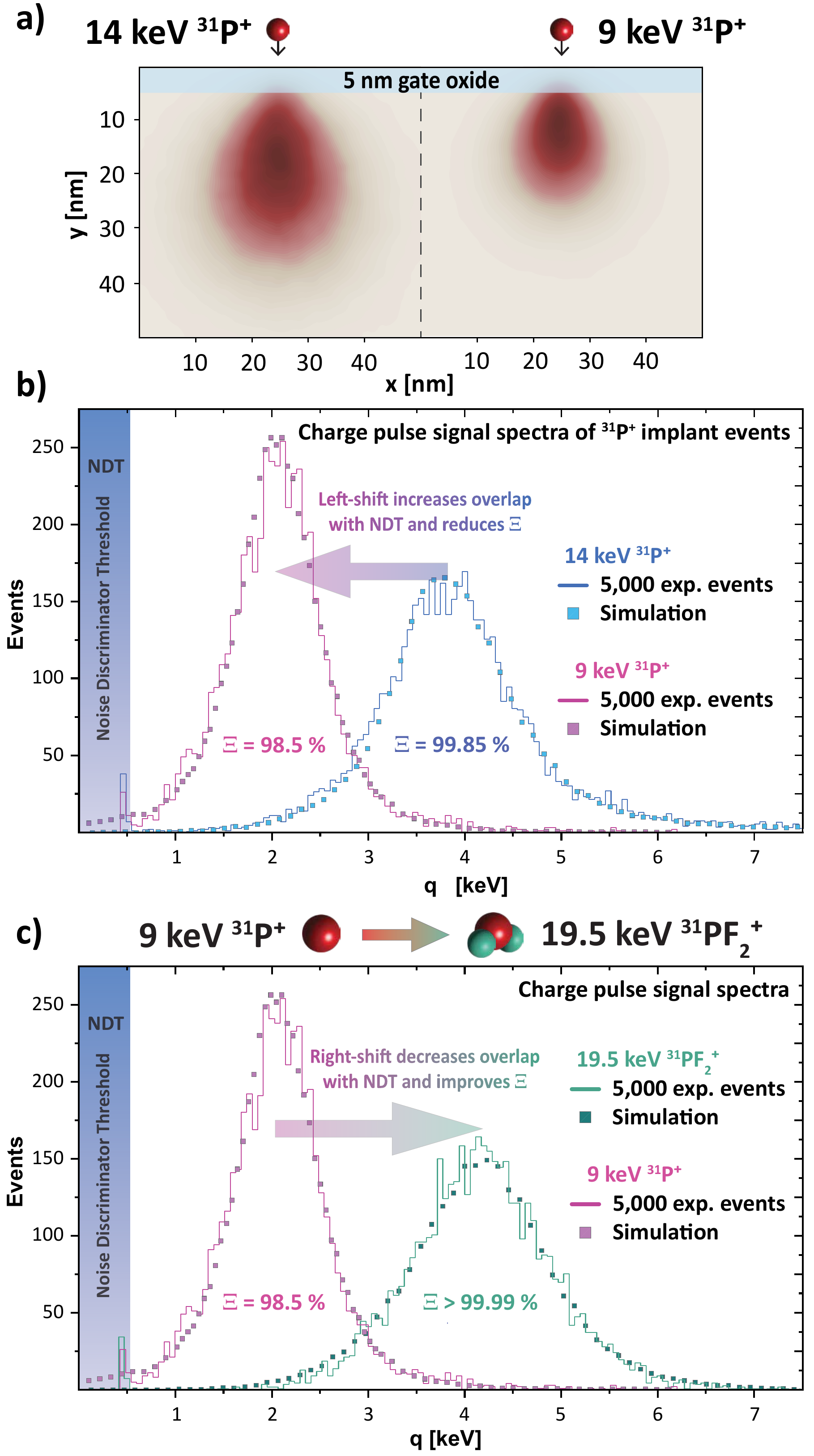}
\caption{\textbf{a)} Comparison of spatial straggling between 14~keV and 9~keV $^{31}$P ions \cite{Ziegler2010}. The reduced ion energy results in an average implantation depth closer to the gate oxide interface and $^\sim$65\% lower placement uncertainty volume from straggling. \textbf{b)} Comparison of the experimental signal spectra from 5,000 $^{31}$P ion implant events for 14~keV and 9~keV. Since 9~keV $^{31}$P ions induce less charge on average, the respective signal spectrum is shifted to lower amplitudes $q$ and increasing overlap with the noise discriminator threshold (NDT). This reduces the single ion detection confidence $\varXi$. \textbf{c)} Reversing this effect by replacing atomic 9~keV $^{31}$P ions with 19.5~keV $^{31}$PF$_2$ molecule ions restores a high single ion detection confidence because of additional charge carriers induced by the F bystander ions.}  
\label{fig:Fig2}
\end{figure}

The implications of the constraints on the ion implantation energy are illustrated in Figure \ref{fig:Fig2}a) that compares $^\sim20$~nm-deep average $^{31}$P donor placement to an optimised donor qubit depth of $^\sim12$~nm, required for optimal access to the donors' electron spin state, and to minimize placement straggle. These depths correspond to kinetic $^{31}$P ion energies of 14~keV and 9~keV respectively. 
 
As expected, lowering the kinetic energy from 14 to 9~keV results in a shift of the signal spectrum $s(q)$ to lower amplitudes $q$, because 9~keV $^{31}$P ions induce on average $^\sim$55\% less charge carriers. As a result, a higher fraction of the spectrum of single ion implantation signals extends into the noise-dominated regime below $q_\mathrm{NDT}$ and is discarded. 

The optimum single-ion detection confidence $\varXi_{\mathrm{opt}}$ for an ion species and implantation energy is obtained by using the simulated signal spectrum $s(q)$ from the Crystal TRIM code \cite{Posselt1992,Jakob2022}, fitted to the experimental data with high precision (see Fig. \ref{fig:Fig2}b)). The calculation yields $\varXi_{\mathrm{opt}}=98.6$\% for 9~keV $^{31}$P ions, meaning that $^\sim$1.4\% of the ions are implanted but not registered which is ten time more than for 14~keV $^{31}$P$^+$ ions.  Molecule ions address this problem as discussed in the next section.

\subsection*{Outward scaling: High placement precision with molecule ions}
\begin{figure*}[t!]
\centering
\includegraphics[width=1.0\linewidth]{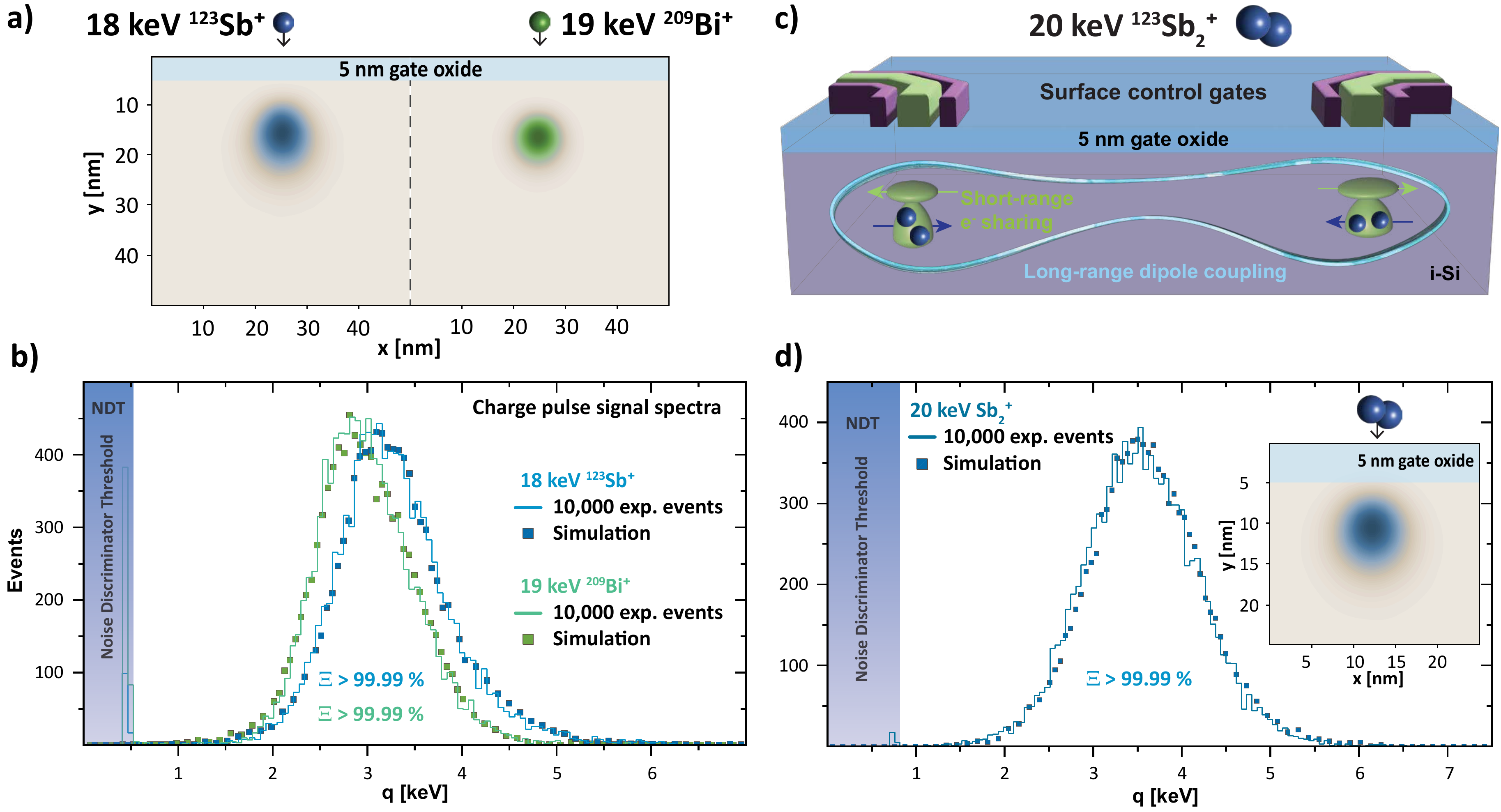}
\caption{\textbf{a)} Spatial straggling of 18~keV $^{123}$Sb and 19~keV $^{209}$Bi ions \cite{Ziegler2010} amounts to $^\sim$5~nm and $^\sim$3~nm, respectively. \textbf{b}) Experimental signal spectra composed of $^\sim$10,000 single ion implant events for both ion species are overlaid with simulated signal spectra. \textbf{c)} Schematic of proposed operation of two distantly spaced $^{123}$Sb donor pairs, where nuclei within the pair are coupled through their shared electron, and distant pairs are coupled by electric dipole interaction. \textbf{d)} Signal spectrum acquired for 10,000 20~keV Sb$_{2}$ molecule ion implant events (acquired at +5~$^\circ$C substrate temperature), overlaid with the simulated spectrum. The contour plot inset shows the spatial distribution of the implanted donor pair due to the combined straggling. The high single ion detection confidence enables scalable engineering of such closely spaced multi-dopant platforms.}
\label{fig:Fig3}
\end{figure*} 

The signal detection hardware employed in this study is optimised for near-room temperature operation (-15...+5~$^\circ$C) and scale-up compatibility. This and the integrability with qubit device fabrication limit leeway for optimising the detector noise performance further. Instead, we adopt the strategy of implanting molecule ions, consisting of $^{31}$P and bystander atoms.

Introducing a molecule ion, where the primary $^{31}$P dopant is supplemented by bystander atoms, modifies the kinetic ion energy to:
\begin{equation}
E_{\mathrm{Ion}} = \frac{M_{\mathrm{p}}+n_{\mathrm{b}} M_{\mathrm{b}}}{2}\,v^2_{\mathrm{p}} \propto q 
\end{equation}
The signal amplitude $q$ increases with kinetic ion energy $E_{\mathrm{Ion}}$ - and consequently with the total ion mass. In turn, adding a number $n_{\mathrm{b}}$ of bystander atoms to the primary implant species $^{31}$P with a desired near-surface placement precision ($v_\mathrm{p}=\mathrm{const.}$), will create additional free electron-hole pairs
and thus increase the detectable signal $q$. This shifts the entire signal spectrum $s(q)$ to higher amplitudes - away from the noise discriminator threshold $q_\mathrm{NDT}$ - and increases the single-ion detection confidence.
. 
A beam of $^{31}$PF$_2$ molecule ions, generated from a $^{31}$PF$_5$ gas source, is utilised in this study. Fluorine is an ideal bystander ion because of its ultra-fast diffusion in silicon, as evidenced via systematic secondary-ion mass spectroscopy (SIMS) analysis \cite{Holmes2023, Ruffell2005}. The rapid thermal anneal (RTA), conducted to activate the implanted $^{31}$P donors, drives practically all fluorine atoms away from the active donor qubit region so that no $^{19}$F-$^{31}$P nuclear spin interactions are observable in a qubit device \cite{Holmes2023}. The signal boost effect of the bystander ions is shown in Figure \ref{fig:Fig2}c), which compares the signal spectra for 5,000 implanted 9~keV $^{31}$P ions versus 19.5~keV $^{31}$PF$_2$ ions. The molecule ion contains the primary $^{31}$P atom - carrying $^\sim$8.7~keV implantation energy - and each $^{19}$F bystander carrying $^\sim$5.4~keV. As expected, the $^{31}$PF$_2$ signal spectrum peak center is characterised by a pronounced shift to higher amplitudes and resides completely above the noise discriminator threshold. Its quantitative assessment yields a single ion detection confidence exceeding 99.99\% (vs 98.6\% for atomic 9~keV $^{31}$P ions). 

\subsection*{Inward scaling: High placement precision for heavy ions}
By employing the same system described already, we show here that two high-nuclear spin donors species, $^{123}$Sb and $^{209}$Bi, can be implanted with high confidence. Again, the kinetic ion energies were chosen for an average placement depth of $\approx$12~nm (Fig. \ref{fig:Fig3}a) below the gate oxide interface.  
The experimental spectra of 10,000 ion impact signal events are shown in Fig. \ref{fig:Fig3}b). The single-ion detection confidence exceeds 99.99\% for both donor species.

The heavy nuclei species $^{123}$Sb and $^{209}$Bi exhibit very small spatial straggling of only 5~nm and 3~nm, respectively (Fig. \ref{fig:Fig3}a). As a result, the fraction of ions that stops inside the 5~nm surface gate oxide is negligible ($\leq$10 per million implants) and back-scattering is absent. Hence, the qubit yield is subject only to the activation yield of the implanted dopants from an appropriate annealing strategy.
%%Hence, the deterministic doping yield concurs with obtained single ion detection confidences. 

\subsection*{Inward-Outward scaling: Heavy donor molecule ions}
\begin{figure*}[t!]
\centering
\includegraphics[width=1.0\linewidth]{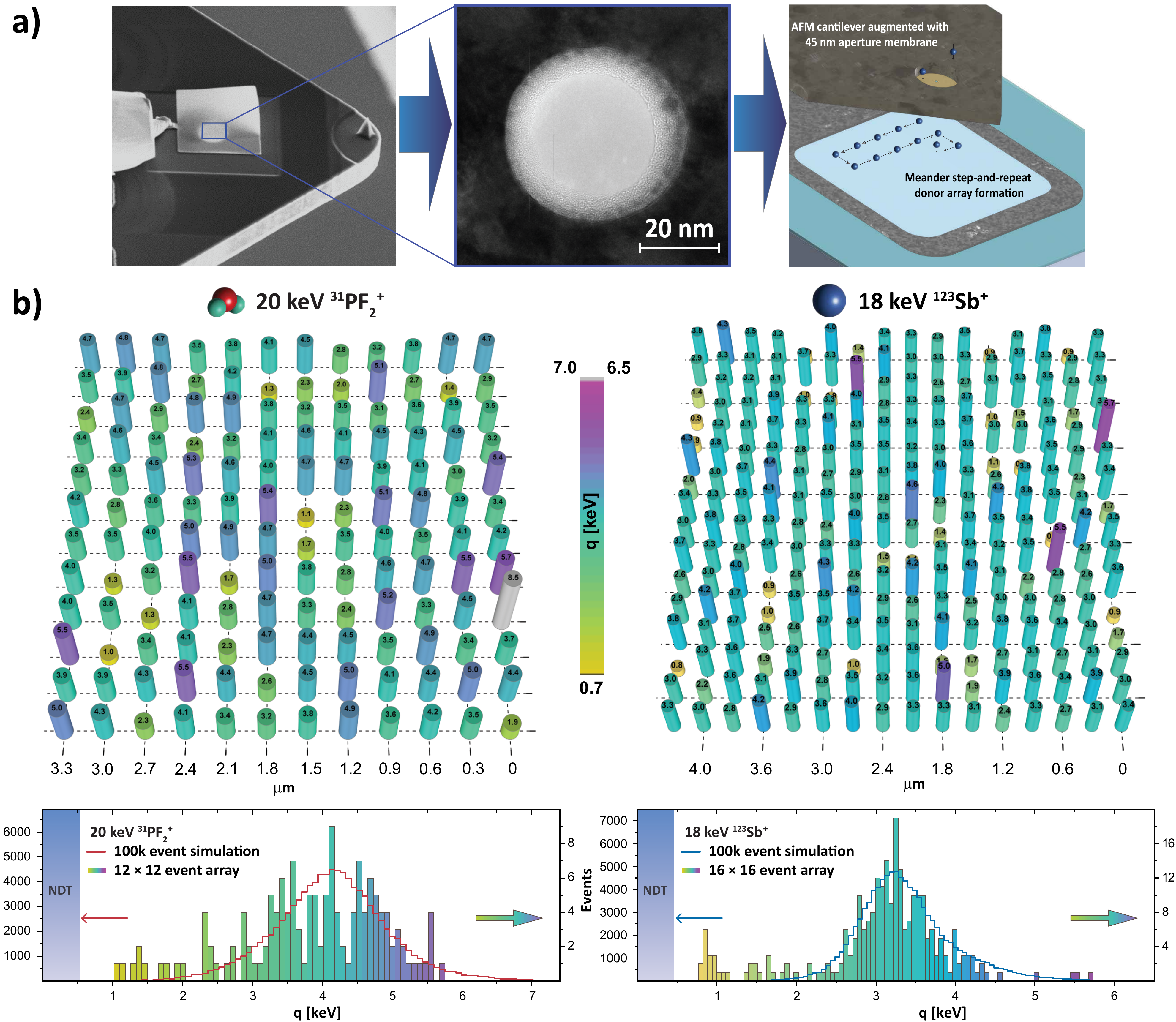}
\caption{\textbf{a)} AFM cantilever configured with a custom-engineered 45~nm nano aperture membrane for donor array fabrication (see Methods).  \textbf{b)} Experimental signal arrays for step-and-repeat implantation of 12$\times$12 (144) $^{31}$PF$_2$ ions at 20~keV and 16$\times$16 (256) $^{123}$Sb ions at 18~keV, together with corresponding signal spectra and simulated spectra for 100,000 implants. The latter are normalised to the total number of the experimental implant events (histogram area) for direct comparison.}
\label{fig:Fig5}
\end{figure*}

For $^{31}$P donor qubits, a significant milestone was the  demonstration of 1- and 2-qubit logic operations with over 99\% fidelity, using a  $^{31}$P donor pair that shared a common hyperfine-coupled electron \cite{Madzik2022} . The pair separation was found to be $\sim 5$~nm apart following a Poissonian implant strategy. However, there is no present technology to deterministically manufacture devices with pairs of $^{31}$P donor qubits at such a close spacing owing to the precision limitations imposed by $^{31}$P ion straggling.\\

This is not the case with Sb$_2$ molecule ions, where our present interest is in molecule ions comprising only the $^{123}$Sb ($I = 7/2$) isotope.  These ions provide a deterministic pathway to create closely-spaced donor pairs. A 20~keV Sb$_2$ molecule ion, implanted at 20~keV energy, dissociates instantaneously upon impact with the substrate, creating a pair of 10~keV Sb fragments; the low spatial straggling at this energy results in an average placement depth about 7~nm below the gate oxide interface, with an average donor spacing of $^\sim$5~nm (Fig. \ref{fig:Fig3}d). This is precisely the required distance for two donors to share an electron, enabling short-range entangling 2-qubit gates mediated by electron rotations \cite{Madzik2022} and the implementation of single-step parity check gates \cite{ustun2023single}. Longer-range entangling gates could then be obtained by exploiting the flip-flop qubit architecture \cite{Tosi2017,Savytskyy2022}, where the gate-induced displacement of the negative electron charge with respect to the positive nuclear charges induces a strong electric dipole, capable of mediating entangling operations between donor-pairs placed 150-500~nm apart (Fig. \ref{fig:Fig3}c).

An experimental 20~keV Sb$_2$ molecule ion signal spectrum, composed of 10,000 single implant events, is illustrated in Fig. \ref{fig:Fig3}d) together with the simulated spectrum. The dataset shows that the single ion detection confidence exceeds 99.99\%, with a deterministic donor-pair doping yield of $^\sim$99.98\%, limited by the $^\sim$0.02\% of 10~keV $^{123}$Sb ions that stop inside the gate oxide \cite{Ziegler2010}. Noteworthy here is that this result was obtained despite a higher substrate operation temperature of +5$^\circ$C (vs -15$^\circ$C for all other), as can be seen from the slightly elevated noise discriminator threshold of $q_\mathrm{NDT}=0.75$~keV. This example of inward-outward scaling can be applied to any practical multi-atomic configuration for various quantum technological applications.

\subsection*{Engineering deterministic donor arrays}
To demonstrate the deterministic ion implantation system \cite{Jakob2022} for engineering ordered near-surface 2D arrays of single donors in silicon, an automated step-and-repeat single ion implantation routine is employed using an AFM cantilever that is configured with a custom-developed 45~nm nano aperture to collimate the ion beam. The signal induced by a single ion implant event triggers the substrate stepping between array implant sites within a pre-defined positioning path (see Methods).   
 
The aperture geometry shown in Fig \ref{fig:Fig5}a) is optimised to reduce ion scattering to about 0.6\% of the incident ions \cite{Raatz2019} (Methods).

For the experimental results presented here, the step pitch between implant sites was 300~nm, commensurate with the flip-flop donor qubit layout, which can theoretically accommodate 150-500~nm inter-donor pitch \cite{Tosi2017}.  Two examples are shown in Fig. \ref{fig:Fig5}b) with 12$\times$12 $^{31}$P and 16$\times$16 $^{123}$Sb single donor arrays. The signal amplitude from each implant site is displayed as a 2D histogram. Also shown are corresponding signal spectra for each array, overlaid with the simulated spectra from 100,000 implant events. This comparison enables the identification of implantation signal events that arise from effects extraneous to the substrate. Indeed, given statistical limitations of 144 and 256 events, respectively, both signal histograms have a satisfactory concordance with the simulated spectra. Hence the majority of single ion implant events is not affected by nano aperture scattering before implanting into the substrate.\\

However, the experimental spectra show a small excess signal clustering within the $[0.5,2]$~keV signal window for both arrays.  The 2D histograms in Fig. \ref{fig:Fig5}b) reveal that these signals are not randomly distributed, but occur mainly in 3 or more adjacent array sites .
 
This indicates surface features of the detector as the primary origin such as localised variations of the charge collection efficiency due to quality variations of the gate oxide or surface debris which decelerates incident ions before they enter the sensitive silicon bulk to induce the signal \cite{Robson2022}. Indeed, the single ion detector substrate utilised here is protected by a PMMA resist film that is removed just prior to the experiment. However, it was found that residues from incomplete removal of the resist resulted in signal attenuation spanning two or three adjacent array sites in isolated areas. Apart from this, the good match between simulated and experimental signal spectra (>2~keV) is quantitatively supported by a $\chi^2$ spectrum comparison test (see Methods). The test identifies the signal distribution of both experiments to lie within a 60\% quantile of respective reference data, which is representative of the fidelity for constructed donor arrays.\\

%******************************************************************************************
%\section*{Conclusion}
This study demonstrates how to transform silicon into a material containing high-fidelity near-surface arrays of single dopants. First, the single ion detection confidence was benchmarked in light of dopant array dimension limitations ('outward' scaling) and demonstrated to exceed 99.99\% for the application-relevant donors $^{31}$P, $^{123}$Sb and $^{209}$Bi. The mass of these donor atoms spans the entire practicable range of the periodic table, thus providing evidence that the method is suitable for any other atomic platform in silicon. Second, for the lightest relevant dopant, $^{31}$P, it was shown how $^{31}$PF$_2$ molecule ions can be exploited for high fidelity deterministic implantation of near-surface arrays, owing to a fluorine bystander-induced signal boost. Finally, this concept has been extended to deterministic Sb$_2$ molecule ion implantation for fabricating closely-spaced donor pairs. This and similar dopant-compound platforms aim to provide many spin qubits per gate and to reduce complexity and footprint of the network infrastructure. Standard tools of the silicon semiconductor industry were adapted here to open new pathways for spin qubit entanglement investigations and device scale-up.

\section*{Methods}
Further details on detector fabrication, charge-sensitive electronics and the ion beamline configuration can be found in \cite{Jakob2022}.\\

\textbf{Selection of molecule ions:} When introducing molecule ions for signal boosting, the bystander atoms should exhibit a comparable or higher nuclei mass than the primary dopant (for sufficient total kinetic ion energy increase and thus boost of the signal $q$) and ideally not interfere with other yield-determining factors of functioning CMOS nano electronic donor qubit devices (e.g. donor activation yield, gate oxide pin hole density etc.). A reasonable choice is the selection of bystander atoms that match the substrate species (i.e. $^{28}$Si) to avoid silicon substrate contamination. Furthermore, the number $n_{\mathrm{b}}$ of molecule bystander atoms should be kept low to reduce ion-induced damage in the silicon lattice and preserving the charge collection efficiency when implanting large dopant arrays (i.e. $\geq$20k dopants). Since the generation of $^{31}$P$^{28}$Si molecule ions in a DC filament ion source typically requires highly corrosive and toxic gases like phosphine and silane, a $^{31}$PF$_2$ molecule ion - generated from less critical PF$_5$ gas - is utilised here as alternative.\\ 
\textbf{Simulation of the signal spectrum:} In a first step the TRIM code \cite{Ziegler1985} is employed in order to simulate the transmission of the incident ion through the thin surface gate oxide on the device. In the case of the molecular ions PF$_2$ and Sb$_2$ it is assumed that these dissociate upon impact and the transmission of the atomic constituents through the oxide is treated separately taking into account their respective kinetic energies. This simulation step yields the kinetic energy and the direction of motion of the transmitted particles and of the recoiled Si and O atoms from the oxide. In the second step, these data are used as inputs for the Crystal-TRIM code \cite{Posselt2003,Posselt2006,Pilz1998} which treats the motion of the aforementioned atoms in the single-crystalline (100) i-Si substrate (see Fig. 1). During this simulation the electronic energy loss per incident atomic or molecular ion is determined. This quantity corresponds to the signal measured by the detector. Finally, detector noise and Fano statistics of e-h pair generation were taken into account and allow the direct comparison with the experimental signal spectrum.\\
Details on Crystal-TRIM program: The code simulates the trajectories of energetic projectiles (in the present work: P, Sb, Bi, F, Si, and O) in single-crystalline Si, in contrast to the TRIM code, which is only applicable to amorphous materials. Therefore, the projectile motion into channeling as well as non-channeling directions can be treated. The former leads to larger values of the electronic energy deposition than the latter. Like TRIM, Crystal-TRIM is based on the binary collision approximation, which assumes that the motion of an energetic projectile may be described by a sequence of binary collisions with target atoms. In this manner, the trajectory of a projectile between two subsequent collisions is approximated by a straight line given by the asymptote to the trajectory of the energetic particle after the first collision. The impact-parameter dependent electronic energy loss during the collision of a projectile with a target atom is treated by a semi-empirical formula, which is similar to the Oen-Robinson model \cite{Posselt2003,Posselt2006,Oen1976}. The formula contains two fit parameter, C$_\lambda$ and C$_\mathrm{el}$, the values used in this work are shown in Table \ref{tab:table1}.\\ 
\begin{table}[b]
\caption{\label{tab:table1} C$_\lambda$ and C$_\mathrm{el}$ parameter used for simulated  signal spectra via fitting to experimentally obtained data with a 16~$\mu$m aperture. 
}
\begin{ruledtabular}
\begin{tabular}{lcccc}
\textrm{Ion \& Energy}&
\textrm{Primary}&
\textrm{Bystander}&
\textrm{O}&
\textrm{Si}\\
\colrule
14~keV P & 0.94, 2.8 & - & 1.0, 2.8 & 1.1, 2.8\\
9~keV P & 0.94, 2.8 & - & 1.0, 2.8 & 1.1, 2.8\\
20~keV PF$_2$ & 0.93, 2.8 & 0.74, 2.8 & 1.2, 2.8 & 1.2, 2.8\\
18~keV Sb & 1.27, 2.8 & - & 0.97, 2.0 & 0.98, 2.0\\
20~keV Sb$_2$ & 1.37, 2.8 & 1.37, 2.8 & 0.97, 2.0 & 0.97, 2.0\\
19~keV Bi & 1.38, 2.8 & - & 1.0, 2.8 & 1.0, 2.8\\
\end{tabular}
\end{ruledtabular}
\end{table}
The parameter C$_\lambda$ describes the average electronic energy loss for non-channeled projectiles. Therefore, C$_\lambda$ determines approximately the energy related to the peak center of the electronic energy loss/charge pulse spectrum per incident ion. The shape of the spectrum is sensitive to the parameter C$_\mathrm{el}$, which influences the channeling of a projectile. Thermal vibrations of lattice atoms affect projectile trajectories, especially for the motion in channels. A simple model is used to take into account this effect \cite{Posselt2003}. In Crystal-TRIM only the motion of the energetic projectiles (incident on (100) Si after transmission through the thin gate oxide) is simulated. However, also Si target atoms set into motion by the collision with the projectiles, and their corresponding collision cascades contribute to the electronic energy deposition. This contribution is described by the experimentally-derived model of Funsten et al. \cite{Funsten2004}, which considers non-negligible self-trapping mechanisms due to a high density of low-energy e-h pairs generated closely around the lattice defect path from energetic ion projectiles. As shown in Fig. \ref{fig:Fig2}c, a satisfactory agreement between present experimental data and those obtained from the Funsten model is found for P projectiles. Hence, also for energetic Si target atoms the use of this model is justified because of similar atomic numbers and masses. On the other hand, the widely used model of Robinson [Ref1,2, see comment in main text] to describe the partition of the energy of a Si target recoil into electronic and nuclear energy deposition does not yield satisfactory results for the electronic energy loss spectrum. At this point, it must be emphasized that the knowledge of the electronic stopping of projectiles with low velocity (e.g. below one tenth of the Bohr velocity) is very limited while these velocities are of particular interest in this work. This lack of data is also a reason that the motion of the Si target recoils is not treated explicitly in present simulations.\\
\begin{figure}[t!]
\centering
\includegraphics[width=1.0\linewidth]{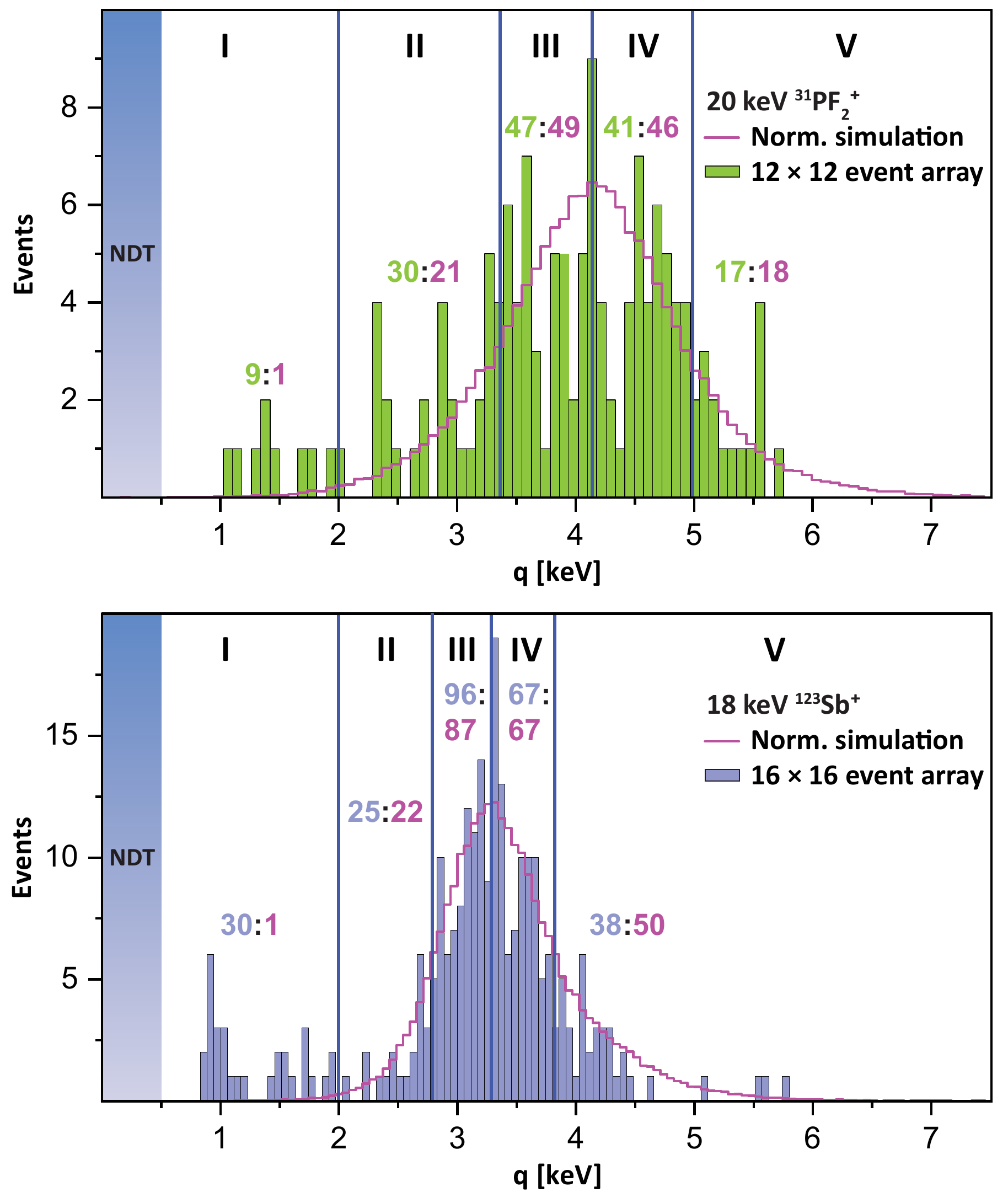}
\caption{Comparison between experimental  signal spectra and re-scaled simulation. The experimental dataset originate from deterministic donor array formation through a stepped 45~nm nano aperture. The simulated spectra are composed of 100,000 events and normalised to the total experimental number of detected events for direct comparison using a $\chi^2$-test. The diagrams visualise class intervals I-V and respective event abundance of experiment vs simulation. For further details see Table \ref{tab:table2}.}
\label{fig:Fig7}
\end{figure}
\textbf{Deterministic donor array formation:} For the automated deterministic dopant array implantation routine (Labview-based protocol), a valid signal event induced by a single ion impact (here any signal pulse above the noise discriminator threshold), triggers a custom-developed ultra-fast blanker circuit output a voltage HI to an electrostatic deflector plate assembly installed above the AFM. The blanker circuit is connected directly to the pre-amplifier to receive the raw analogue signal pulse (reducing response time and monitored/controlled via PC with a separate integrated circuit). The biased deflector diverts the pre-collimated ion beam (20~$\mu$m) away from the cantilever nano aperture ($\leq$100~ns response time to suppress multiple ion implants per site). A TTL pulse sent by the PX5 digital pulse processor to the AFM stage controller advises to proceed with sample re-positioning to the next implant site within a pre-defined meander-like stepping path. Each step cycle concludes with a TTL pulse by control PC to the beam blanker to trigger a voltage LO at its output, re-blanking the ion beam back onto the cantilever nano aperture and awaiting the next single ion implant event.\\
\textbf{Donor array analysis:} A $\chi^2$ test was employed to assess the compatibility of the recorded spectral signal distributions for the implanted $^{31}$PF$_2$ and $^{123}$Sb arrays with the respective computed spectra. As the recorded events in the energy range from 0.5~keV to 2.0~keV are caused by energy losses in the PMMA residuals on top of the gate oxide surface, they are not expected to fall within the computed spectra and were thus excluded from the analysis. Due to the low overall number of recorded events (144 for $^{31}$P$_2$ and 256 for $^{123}$Sb), five  event classes I-V were formed for each spectrum peak with the following boundary definitions (see Fig. \ref{fig:Fig7} and Table \ref{tab:table2}): 2~keV, left FWHM of peak center, peak maximum, right FWHM of peak center, and high-energy tail end of spectrum. The simulated reference signal spectra, composed of 100,000 implant events, were normalized to the total number of experimentally recorded events above 2~keV, i.e. 135 events for $^{31}$PF$_2$ and 226 events for $^{123}$Sb ion implantation.\\
\begin{table}[b]
\caption{\label{tab:table2} Tabulated results of $\chi^2$ distribution comparison tests for  spectra from 18~keV $^{123}$Sb and 20~keV $^{31}$PF$_2$ ions, implanted through a 45~nm nano aperture (see Fig. \ref{fig:Fig7}). The simulated spectra are composed of 100,000 implant events and serve as reference. The p-Quantile is a measure of the (mis)match between both signal spectra (event classes) and describes the covered area of a Poisson probability distribution $P(x)$. A p-Quantile value of 0.6 ($\approx1\times\sigma$) corresponds to the area coverage at FWHM of the distribution and can be interpreted as reasonable probability that test class of events reflects the event abundance of the reference class. On the contrary, p-Quantile values approaching 1 mean that the event class under test falls into the far outer tail of the Poisson distribution and are increasingly unlikely to reflect the event abundance of the reference class. Class I is excluded for the summation, as the number of excess events therein is caused by PMMA residuals on the detector surface, which decelerate impacting ions before they can induce the signal $q$ in the device.}
\begin{ruledtabular}
\begin{tabular}{lccccr}
\textrm{Ion}&
\textrm{Class}&
\textrm{Exp.}&
\textrm{Ref.}&
\textrm{$\chi^2$}&
\textrm{p-Quant.}\\
\colrule
\textbf{$^{31}$PF$_2$} & I & 9 & 1.1 & 58 & $>$0.99\\
              & II & 30 & 21.1 & 3.75 & 0.95\\
              & III & 47 & 48.5 & 0.05 & 0.2\\
              & IV & 41 & 46 & 0.54 & 0.5\\
              & V & 17 & 18.3 & 0.09 & 0.25\\              
              & \textbf{II-V} & \textbf{135} & \textbf{135} & \textbf{4.43} & \textbf{0.6}\\\\
\textbf{$^{123}$Sb} & I & 30 & 0.9 & 971 & $>$0.99\\
              & II & 25 & 21.8 & 0.49 & 0.5\\
              & III & 96 & 87 & 0.93 & 0.65\\
              & IV & 67 & 66.6 & $<$0.01 & 0.05\\
              & V & 38 & 50.6 & 3.14 & 0.92\\
              & \textbf{II-V} & \textbf{226} & \textbf{226} & \textbf{4.56} & \textbf{0.6}\\
\end{tabular}
\end{ruledtabular}
\end{table}
\textbf{Ion aperture scattering analysis:} To simulate the scattering of ions through the nano aperture, the iradina code was employed \cite{Borschel2011}. This program bases on the TRIM software \cite{Ziegler2010} and simulates the interactions of ions with a target specimen using a Monte-Carlo transport algorithm. Instead of a planar geometry, any arbitrary nano structure can be defined using a three-dimensional grid of cells. Here, a 100$\times$100$\times$100 simulation volume with a cell size of \SI{1}{nm} was defined. The ion beam propagates in the $\hat{x}$ direction, and encounters the target that spans the $y$-$z$ plane. The target structure was created using CAD software: a plane in the range $x =$ [0, 15) was first defined, corresponding to the surface Au coating. The Si$_3$N$_4$ body was then defined with another plane spanning $x = $ [15, 50). Finally, the aperture itself was created by overwriting parts of the existing structure with a tapered cone comprised of vacuum. One end was centred at the origin with a diameter \SI{45}{nm}, while the other was centred at (50,0,0) with diameter \SI{45.87}{nm}, giving the \SI{0.5}{^\circ} sidewall angle. A noise component of \SI{1}{nm} (r.m.s) was added to the cone sidewalls to achieve the desired roughness. To mimic the experimental conditions, the beam was defined to enter the aperture concentrically and with uniform lateral intensity distribution. To minimise computation time, the beam diameter was limited to twice that of the aperture, because ions that undergo small-angle scattering with the aperture edges are of primary interest rather than internal collision cascades in the aperture material. It should be noted that aperture material modification due to surface sputtering and implantation of incident ions is negligible for employed ion fluences and exposure times, which justifies the utilisation of this code.\\ 
In general, the fraction of incident ions affected by aperture edge scattering increases inevitably with aperture down-scaling to the nano meter scale, as the ratio between cross-sectional aperture area and its perimeter edge reduces. To estimate the degree of this effect, the scattered ion fraction is computed \cite{Borschel2011} for the 18~keV $^{123}$Sb experiment with the nano aperture employed in the experiments. Figure \ref{fig:Fig6}b) illustrates the kinetic energy distribution of the ions after passing the modelled aperture (Configuration I) as well as a second hypothetical nano aperture Configuration II. The latter is an unfavourable geometric configuration for comparison and is characterised by a 4$^\circ$ aperture channel taper angle and 2$^\circ$ angular misalignment with respect to the incident ion beam. As can be seen, a total portion of about 0.6\% and 4.4\% of the passing $^{123}$Sb ions are affected by scattering for the respective aperture configurations. These distributions serve as an input set for simulating the affected signal spectra, shown in Fig. \ref{fig:Fig6}c). It is immediately apparent, that not even the worst-case aperture Configuration II is capable of explaining the experimentally observed signal distribution and abundance within $[0,2]$~keV. For the experimentally relevant nano aperture Configuration I, the ion scatter statistics are relaxed and not further addressed here. As for the hypothetical aperture Configuration II; out of 4.4\% scattered ions, the majority (3.4\%) is problematic for donor array engineering, as these ions induce signals that lie within the main signal spectrum peak and are not reliably distinguishable from non-scattered ions. These events represent ions that are likely implanted outside the aperture confinement window due to a significant angular alteration of their momentum vector up to 10$^\circ$. Hence these events are likely not within reach of subsequently implemented surface contol gates and constitute qubit loss faults. Another 0.4\% are scattered ions that can be detected and distinguished from regular events. These events are less problematic as a forced second implant event at the same site is likely to result in a regular single donor. Finally, about 0.6\% are non-detectable ion scatter events, as they fall below the noise discriminator threshold. These are of no concern since the low-energetic $^{123}$Sb ions stop entirely within the gate oxide and remain electrically passive. In practice, the nano aperture dwells at the site until an implant event is detected.\\
Based on these results, ion scattering at the aperture edge can be excluded as primary source for observed discrepancy between simulated reference signal spectra and experimental signal histograms for nano aperture single donor array formation experiments. In a wider sense, the simulated results indicate that a solid state nano aperture can be designed to meet the demands for engineering at least small and meso-scaled deterministic dopant arrays suitable for multi-qubit entanglement studies. Moreover, these findings for employed solid state nano apertures do not constitute a general scalability limitation for the employed single ion detectors, since the latter can be combined with a scatter-free focused ion beam (FIB) approach \cite{Robson2022}.\\ 
It is worth noting that implantation signal events from forward recoiled atoms sputtered from the aperture itself (Si, N, Au) are extremely rare due to the optimised aperture design (totalling $\leq$0.01\%) and are not further addressed here.
\begin{figure}[t!]
\centering
\includegraphics[width=1.0\linewidth]{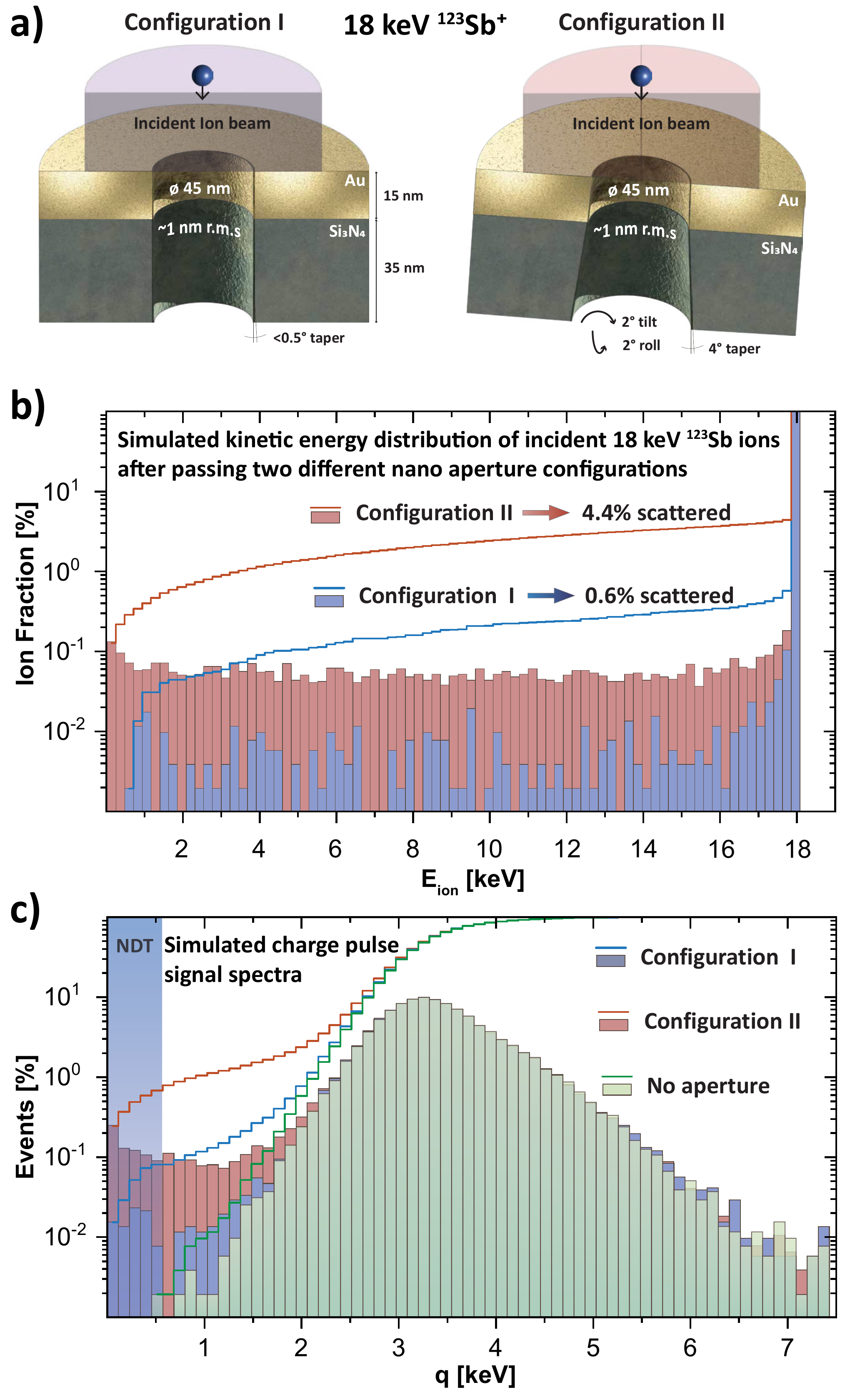}
\caption{\textbf{a)} Schematic of two different geometric nano aperture configurations exposed to an incident beam of 18~keV $^{123}$Sb ions. \textbf{b)} Simulated kinetic ion energy distributions of $^\sim$50,000 (i.e. 100\%) ions after passing through two nano aperture configurations. \text{c)} Corresponding simulated signal spectra, plotted on a log scale for improved visibility, of the signal event percentage that originates from scattered ions. Cumulative distribution line plots are provided in \textbf{b)} and \textbf{c)}.}
\label{fig:Fig6}
\end{figure}

\begin{acknowledgments}
The research at the University of Melbourne and UNSW was funded by the Australian Research Council Centre of Excellence for Quantum Computation and Communication Technology (Grant No. CE170100012) and the US Army Research Office (Contracts No. W911NF-17-1-0200 and W911NF-23-1-0113). We acknowledge a grant from the University of Melbourne Research and Infrastructure Fund (RIF) and use of the facilities of the Australian National Fabrication Facility (ANFF) at the Melbourne Centre for Nanofabrication (MCN) and at UNSW.  The authors acknowledge access to the NCRIS funded Heavy Ion Accelerator Capability at the University of Melbourne. S.G.R and H.R.F. acknowledge the support of an Australian Government Research Training Program Scholarship.  A.M. Jakob acknowledges an Australia–Germany Joint Research Cooperation Scheme (UA-DAAD) travel scholarship that supported collaboration with partner institutions in Germany. We are grateful to D. McCulloch of the RMIT Microscopy and Microanalysis Facility for use of SEM/FIB and TEM equipment. The views and conclusions contained in this document are those of the authors and should not be interpreted as representing the official policies, either expressed or implied, of the ARO or the US Government. The US Government is authorized to reproduce and distribute reprints for government purposes notwithstanding any copyright notation herein.
\end{acknowledgments}

\section*{Author contributions statement}
A.M. Jakob conceived and constructed the detector preamplifier electronics and the integration of the ion beam line with the IBIC-AFM hybrid nano scanner. A.M. Jakob conceived and conducted the deterministic implantation experiments with assistance from S.G. Robson. A.M. Jakob developed single ion detectors and respective fabrication protocol. H.R. Firgau, V. Mourik and V. Schmitt fabricated single ion detector substrates. S.G. Robson conducted C-V/I-V analysis for iterative detector performance improvement. S.G. Robson developed the Labview-based step-and-repeat code with conceptual input from A.M. Jakob for the signal-feedback-controlled AFM scanner movement and ion beam blanking for deterministic donor array formation. S.G. Robson conducted systematic Iradina simulations on nano aperture geometries yielding reduced ion scattering. S.G. Robson conceived and performed Ga-FIB and HIM-based ion nano aperture fabrication and conducted SEM analysis. E. Mayes conducted TEM analysis with assistance from S.G. Robson. D. Holmes developed the design and fabrication protocol for implementing silicon CMOS nano electronic devices to single ion detector substrates. D. Spemann and A.M. Jakob developed theoretical groundwork on the detection confidence, signal spectra and donor array analysis. M. Posselt developed a dedicated Crystal-TRIM code for signal spectra simulations. A.M. Jakob and M. Posselt conducted Crystal-TRIM computations and analysis. A. Morello provided the design of the flip-flop qubit architecture, theoretical input on feasibility and device design constraints and supervised the research at UNSW. D.N. Jamieson and A.M. Jakob conceived the deterministic ion implantation process and the implementation of the apparatus. The manuscript was written by A.M. Jakob, D.N Jamieson and A. Morello with contributions from the co-authors.

\section*{Additional information}
The authors declare no competing interests.

\bibliography{apssamp}

\end{document}